\documentstyle[twoside,fleqn,espcrc2,epsf]{article}

\newcommand{\AmS}{{\protect\the\textfont2
  A\kern-.1667em\lower.5ex\hbox{M}\kern-.125emS}}
\newcommand{\GeV}{\mbox{GeV}}
\newcommand{\MeV}{\mbox{MeV}}

\title{
\hfill\begin{minipage}{0pt}\scriptsize\vspace*{-1.5cm} \begin{tabbing}
\hspace*{\fill} UTCCP-P-49 \\ 
\end{tabbing} 
\end{minipage}\\[-8pt]
Heavy Quark Physics in $N_f=2$ QCD\thanks{talk presented by H.P.~Shanahan}}

\author{
CP-PACS Collaboration: \\
\vspace{1mm}
S.~Aoki\rlap,\address{Institute of Physics,
University of Tsukuba, Tsukuba 305-8571, Japan}
R.~Burkhalter\rlap,$^{\rm a,}$\address{Center for Computational Physics, 
University of Tsukuba, Tsukuba 305-8577, Japan}
S.~Ejiri\rlap,$^{\rm b}$
M.~Fukugita\rlap,\address{Institute for Cosmic Ray Research, 
University of Tokyo, Tanashi 188-8502, Japan}
S.~Hashimoto\rlap,\address{High Energy Accelerator Research Organization 
(KEK), Tsukuba 305-0801, Japan}
Y.~Iwasaki\rlap,$^{\rm a,b}$
K.~Kanaya\rlap,$^{\rm a,b}$
T.~Kaneko\rlap,$^{\rm b}$
Y.~Kuramashi\rlap,$^{\rm d}$
K.~Nagai\rlap,$^{\rm b}$
M.~Okawa\rlap,$^{\rm d}$
H.~P.~Shanahan\rlap,$^{\rm b}$
A.~Ukawa$^{\rm a,b}$
and
T.~Yoshi\'e$^{\rm a,b}$
}

\begin{document}

\begin{abstract}
We present a preliminary analysis of the heavy-heavy spectrum and 
heavy-light decay constants in full QCD, using a tadpole-improved SW 
quark action and an RG-improved gauge action on a  $16^3 \times 32$ 
lattice with four sea quark masses corresponding to 
$m_\pi/m_\rho\approx 0.8, 0.75, 0.7, 0.6$  and 
$a^{-1} \approx 1.3\, \GeV$. We focus
particularly on the effect of sea quarks on these observables. 
\end{abstract}

\maketitle

\section{Introduction}
\vspace{-4pt}
Despite the difficulties that heavy quarks present for study on the lattice, 
in two areas at least, they provide an excellent laboratory for examining 
effects of sea quarks, namely the hyperfine splitting of heavy-onia and 
heavy-light decay constants. 


The calculation of the hyperfine splitting has been extensively pursued 
in quenched QCD\cite{draper}. 
In the charm sector, one can compare results with the 
experimental value 
$M_{J/\psi} - M_{\eta\scriptscriptstyle{c}} = 118(2)\MeV$. 
Relativistic actions underestimate this splitting, 
giving a result of $70$-$80 \, \MeV$.
Non-relativistic actions yield similar values. 
This difference has been argued to arise from quenching, 
due to a smaller value of the strong coupling constant at short
distances in quenched QCD\cite{elkhadra}.  
Since the strong coupling constant, as estimated from the Coulomb term 
of the static potential, has been seen to increase by about 10\% 
by the introduction of sea quarks\cite{SESAM,kaneko}, one may hope to 
observe an effect also in the hyperfine splitting. 

A similar reasoning suggests that heavy-light decay
constants are underestimated without sea quarks. 
Recent results from the MILC \cite{milc-prl} and NRQCD \cite{nrqcd-fB}  
Collaborations do show that the decay constants increase with decreasing 
sea quark mass, although an accurate measurement still has to be carried out.

In this report we present preliminary heavy quark results calculated on 
two-flavor full QCD configurations being generated by the CP-PACS 
Collaboration\cite{ruedi}, with emphasis on search for sea quark effects. 
\vspace{-8pt}
\begin{table}[t]
\setlength{\tabcolsep}{.4pc}
\newlength{\digitwidth} \settowidth{\digitwidth}{\rm 0}
\catcode`?=\active \def?{\kern\digitwidth}
\caption{Run parameters for this simulation with eventual final statistics 
in parentheses. }
\label{tab:run-params} 
\begin{tabular}{cccc}\hline
$K_{sea}$ & $m_{PS}/m_V$ & $a^{-1}_\sigma [\GeV]$ & $N_{cfg}$ (HH,HL)   
\\ \hline
0.1375 & 0.8048(9) & 0.9653(65)  & 179, 322 (700) \\ 
0.1390 & 0.751(1)  & 1.0205(80)  & 215, 311 (700) \\ 
0.1400 & 0.688(1)  & 1.0889(72)  & 270, 229 (700) \\ 
0.1410 & 0.586(3)  & 1.1612(87)  & 169, 224 (500) \\ \hline
\end{tabular}
\vspace{-10mm}
\end{table}
\section{Computational Details}
\vspace{-3pt}
Our analysis is carried out on a $16^3 \times 32$ lattice
for four sea quark masses corresponding to $m_{PS}/m_V\approx 0.8$-$0.6$. 
Configurations are generated with an RG-improved 
gauge action at $\beta=1.95$ and the SW quark action with 
the clover coefficient $c_{sw}=P^{-3/4}=1.53$ where $P=1-0.8412/\beta$ is the 
one-loop value of plaquette. 
The lattice spacing,  set either by string tension or
$m_\rho$, 
takes the value $a^{-1}\approx 1.3 \, \GeV$. 
Parameters relevant for the present study 
are listed in Table~\ref{tab:run-params}.

Hadron correlators are computed using local and exponentially 
smeared sources in Coulomb gauge. 
For study of the heavy-heavy spectrum we choose values of $K$ 
corresponding to $am_Q=$ 0.68, 0.75, 0.83 and 0.9 in the naive pole mass 
definition $am_Q=(1/K-1/K_c)/2$.
In conjunction with the above estimate for $a^{-1}$, 
our present study therefore explores the 
region of charm quark. 

For the heavy-light decay constants a set of lighter heavy quark  masses 
are chosen, determined from a plot of the kinetic mass of heavy-light meson 
against $1/K$. The light quark has a mass tuned to approximately 
the strange quark mass as estimated from 
$m^{PS}_{\overline{s}s}/m_\phi = 0.688$ according to 
the Gell-Mann-Okubo formula. 

\begin{figure}[tb]
\centerline{\epsfysize=6cm \epsfbox{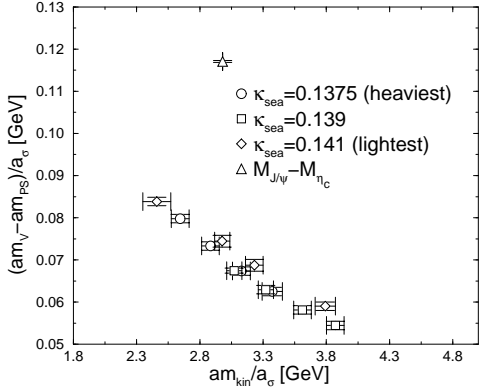}}
\vspace{-30pt}
\caption{Heavy hyperfine splitting using kinetic mass definition. Scale is 
set by string tension measured for each sea quark mass.}
\label{fig:split-kin}
\vspace{-10pt}
\end{figure}

\section{Results}
\subsection{Hyperfine splitting}

We calculate the hyperfine splitting from the difference of pole masses. 
With the use of the SW action lattice discretization effects associated 
with large quark masses are expected to be small for this quantity. 
The pole mass itself, however, does suffer from large discretization 
effects, causing problems in calculating the ground state mass. 
Several possible schemes for measuring the ground state mass 
have been proposed. We compare two :\\
(i) the kinetic mass, derived from the dispersion relation,\\
(ii) an ``HQET'' mass, suggested by 
Bernard {\it et al.}\cite{JLQCD}, defined as 
\begin{equation}
am^{PS}_{HQET} =  am^{PS}_{pole} 
+   N_Q ( am^{Q}_{kin} - am^{Q}_{pole})  ,
\end{equation}
with $N_Q$ the number of heavy quarks. We use tree-level 
quark mass definitions tadpole-improved by a factor $u_0=P^{1/4}$. 

\begin{figure}[tb]
\centerline{\epsfysize=6cm \epsfbox{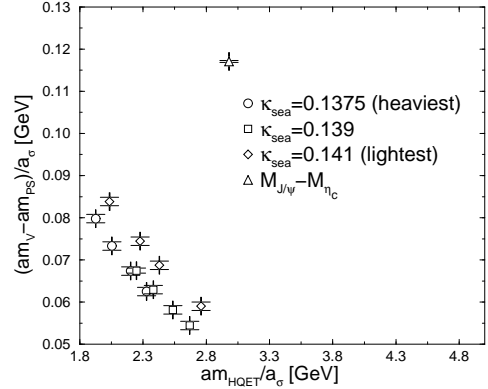}}
\vspace{-30pt}
\caption{Heavy hyperfine splitting using HQET mass definition.}
\label{fig:split-hqet}
\vspace{-10pt}
\end{figure}

We show our results in Figs.~\ref{fig:split-kin} and \ref{fig:split-hqet}. 
The scale is set by the string tension determined for each sea quark 
mass 
to absorb change of scale.  As can be seen, 
the two definitions draw different conclusions for the hyperfine splitting. 
The kinetic mass gives results which indicate an almost negligible change 
between the heaviest sea quark mass ($\approx 3m_{strange}$)
and the lightest ($\approx m_{strange}/2$).  The values 
are roughly consistent with those obtained  from quenched studies. 
Using the HQET mass definition there is a significant difference 
between the heaviest and lightest sea quark masses, but the pseudoscalar 
masses are now much smaller, indicating that at the charm quark mass, 
the hyperfine splitting is smaller, approximately $ 55 \; \MeV$.
\begin{center}
\begin{figure}[tb]
\centerline{\epsfysize=6cm \epsfbox{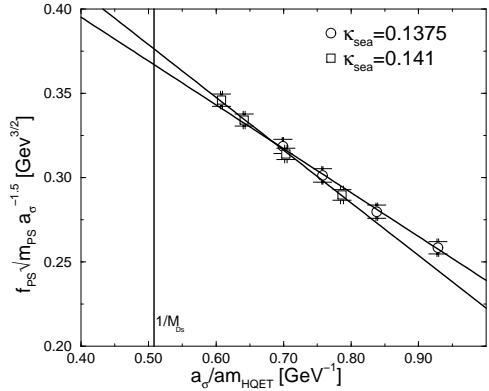}}
\vspace{-30pt}
\caption{$f_{PS}\sqrt{m_{PS}}$  versus $1/m_{PS}$ with a valence strange quark. For clarity 
only the heaviest and lightest sea quark masses and their central fits 
are shown. }
\label{fig:frootm}
\vspace{-20pt}
\end{figure}
\end{center}

\subsection{Heavy-light decay constants}

In Fig.~\ref{fig:frootm} we plot $f_{PS}\sqrt{M_{PS}}$, using the 
KLM normalization for quark fields. The axial vector renormalization
fatcor is taken from Ref.~\cite{taniguchi} calculated for massless quark 
at one-loop perturbation theory.   
As the heavy quark  masses taken are smaller than those for the hyperfine
splitting measurements, the  difference between the two  mass definitions 
is negligible.  We adopt the HQET mass definition which has smaller 
statistical error.  The scale is set by the string tension as before. 

The results for the heaviest and lightest sea quark masses 
are quite similar in magnitude. There is a slight upward change in the 
slope, however, for lighter sea quark.  
An extrapolation to obtain $f_{D\scriptscriptstyle{s}}$, for which 
we adopt an expansion linear in $1/m_{PS}$, yields a 
shift between the heaviest and lightest sea quark masses of 2--3 \%. 

\section{Conclusions}

At present the ambiguity that exists in the ground state meson mass precludes 
any quantitative statement about sea quark effects in the hyperfine 
splitting. We emphasize that this is not symptomatic of the dynamical 
configurations employed but is a general feature of coarse lattice spacings. 
This problem will be alleviated in the set of configurations with a finer 
lattice spacing $a^{-1}\sim \, 2.5 \, \GeV$ generated by CP-PACS, which we plan to 
analyze next. 
Another possible avenue of approach would be to reduce discretization 
effects through 
introduction of further terms in the SW action as suggested in 
Refs.~\cite{tim,fermi}. 

The heavy-light decay constants around the charm mass region examined here 
show only a slight shift upwards as the sea quark mass is reduced. 
Quenched chiral perturbation theory\cite{booth,sharpe}
suggests, however,
that 
the shift increases with the heavy quark mass, and may 
reach a sizable level in the region of the b quark.
Studies with very heavy (or static) quarks may then provide the best region for 
understanding the systematic error due to quenching.

\vspace{2mm}

This work is supported in part by the Grants-in-Aid
of Ministry of Education
(Nos. 08640404, 09304029, 10640246, 10640248, 10740107).
SE and KN are JSPS Research Fellows. HPS is supported
by JSPS Research for Future Program.

\end{document}